\documentclass[conference]{IEEEtran}
\IEEEoverridecommandlockouts
\usepackage{cite}
\usepackage{amsmath,amssymb,amsfonts}
\usepackage{algorithmic}
\usepackage{graphicx}
\usepackage{textcomp}
\usepackage{xcolor}
\usepackage{float}
\usepackage[ruled,vlined]{algorithm2e}
\usepackage{url}
\usepackage{multirow}
\usepackage{makecell}
\usepackage[hidelinks]{hyperref}
\def\BibTeX{{\rm B\kern-.05em{\sc i\kern-.025em b}\kern-.08em
    T\kern-.1667em\lower.7ex\hbox{E}\kern-.125emX}}

\begin{document}

\title{Adaptive State Estimation Under Topological Uncertainty in Unobservable Primary Distribution Systems Using Strategically Placed Sensors
\\
\thanks{This work was supported by the U.S. Department of Energy (DOE) grant DE-OE0000983.

The views expressed herein do not necessarily represent the views of the U.S. Department of Energy or the United States Government.}}
\author{
    \IEEEauthorblockN{Farah Elsherif\textsuperscript{1}, Behrouz Azimian\textsuperscript{2}, Anamitra Pal\textsuperscript{1}}
    \IEEEauthorblockA{\textsuperscript{1}\text{School of Electrical, Computer, and Energy Engineering}, \textit{Arizona State University}, {Tempe, AZ}}
\IEEEauthorblockA{\textsuperscript{2}\text{GE Vernova}, \textit{Bothell}, WA\\
    felsheri@asu.edu, behrouz.azimian@gevernova.com, anamitra.pal@asu.edu}}

\maketitle

\begin{abstract}
The rapid integration of distributed energy resources is fundamentally altering power flow patterns in primary distribution networks and intensifying operational uncertainty. 
These problems are further compounded by lack of real-time situational awareness and frequent topology changes.
To address these problems, this paper
proposes an integrated deep learning framework for simultaneous topology identification (TI) and distribution system state estimation (DSSE) in real-time unobservable primary distribution networks instrumented by a minimal set of synchronized measurement devices (SMDs). A correlation-driven SMD placement algorithm is introduced first that jointly satisfies TI accuracy and DSSE performance requirements
by exploiting temporal 
and spatial correlations in nodal voltage measurements. 
A \textit{dual} deep neural network (DNN)-based DSSE model is developed next to estimate three-phase voltage magnitudes and angles 
across diverse operating conditions.
To extend the framework beyond the base topology, fine-tuning-based transfer learning is employed to adapt the DSSE model to
reconfigured topologies using limited retraining data. The framework is validated under both Gaussian and non-Gaussian measurement noise and benchmarked against a conventional estimation approach and a single DNN model.
\end{abstract}


\section{Introduction}

The rapid integration of distributed energy resources (DERs)
is fundamentally reshaping power flow patterns in distribution networks. High DER penetration introduces voltage fluctuations, reverse power flows, and localized violations that can propagate deep into the feeder and remain entirely undetected without adequate real-time monitoring. Achieving real-time situational awareness is therefore a prerequisite for reliable and secure distribution grid
operation and control
\cite{iea2023,dalal2024}.

However, the \textit{primary} circuit of the distribution system is critically under-instrumented in the context of real-time situational awareness. 
Smart meters, which offer spatially dense coverage and report at 15-minute to hourly intervals, have latencies of several hours; they are also placed on the \textit{secondary} circuits.
SCADA systems provide faster updates but their observability is effectively confined to the feeder head \cite{azimian2025}. Synchronized measurement devices (SMDs), such as
phasor measurement units (PMUs) and micro-PMUs ($\mu$PMUs), 
provide millisecond-resolution time-synchronized measurements 
for
distribution system state estimation (DSSE) 
\cite{naspi2020}. However, the
high deployment cost of SMDs keeps practical installations extremely sparse. Field measurements also confirm that SMD noise is better characterized by non-Gaussian models than the Gaussian assumption adopted in conventional estimators \cite{azimian2020}.

In this regard, deep neural network (DNN)-based state estimators have been developed to address both unobservability \cite{mestav2019} as well as non-Gaussian noise \cite{azimian2022}.
Particularly, \cite{azimian2022} went a step further to not only present a unified DNN framework for simultaneous three-phase unbalanced DSSE and topology identification (TI) using sparse SMD measurements, but also created an SMD placement scheme exclusively for DNNs, demonstrating that DNN-based estimators achieve accurate joint estimation with far fewer sensors than conventional estimators.
However,
the sensor placement in \cite{azimian2022} was derived from spatial correlation analysis at a single hour,
implying that it could not account for temporal variability of voltages 
occurring across different hours of the day.
Moreover, the voltage magnitude and angle in \cite{azimian2022} were estimated by a single 
unified 
DNN, which we demonstrate produces inferior performance compared to
dedicated models for each quantity.

This paper 
proposes a 
comprehensive framework for deep learning-based time-synchronized TI and three-phase unbalanced DSSE to ensure reliable operation
across all hours of the day
in real-time unobservable primary distribution networks. The specific contributions are:
\begin{enumerate}
    \item A \textit{correlation-driven SMD placement algorithm} that jointly optimizes sensor locations for both TI and DSSE by incorporating 
    temporal variability and 
    spatial correlation 
    of all nodal voltage phasors.
    \item \textit{Dual DNN-based DSSE models} that independently estimate three-phase voltage magnitude and angles, demonstrating superior accuracy over a single 
    unified 
    model across different hours of the day
    and under both Gaussian and non-Gaussian noise.
    \item A \textit{fine tuning (FT)-based transfer learning (TL)} strategy that adapts the base-topology DSSE model to reconfigured network topologies using limited retraining data, triggered automatically by the DNN-based TI model upon topology change detection.
\end{enumerate}

\begin{figure}[ht]
    \centering
    
    \vspace{-0.5em}    
    \includegraphics[width=0.485\textwidth]{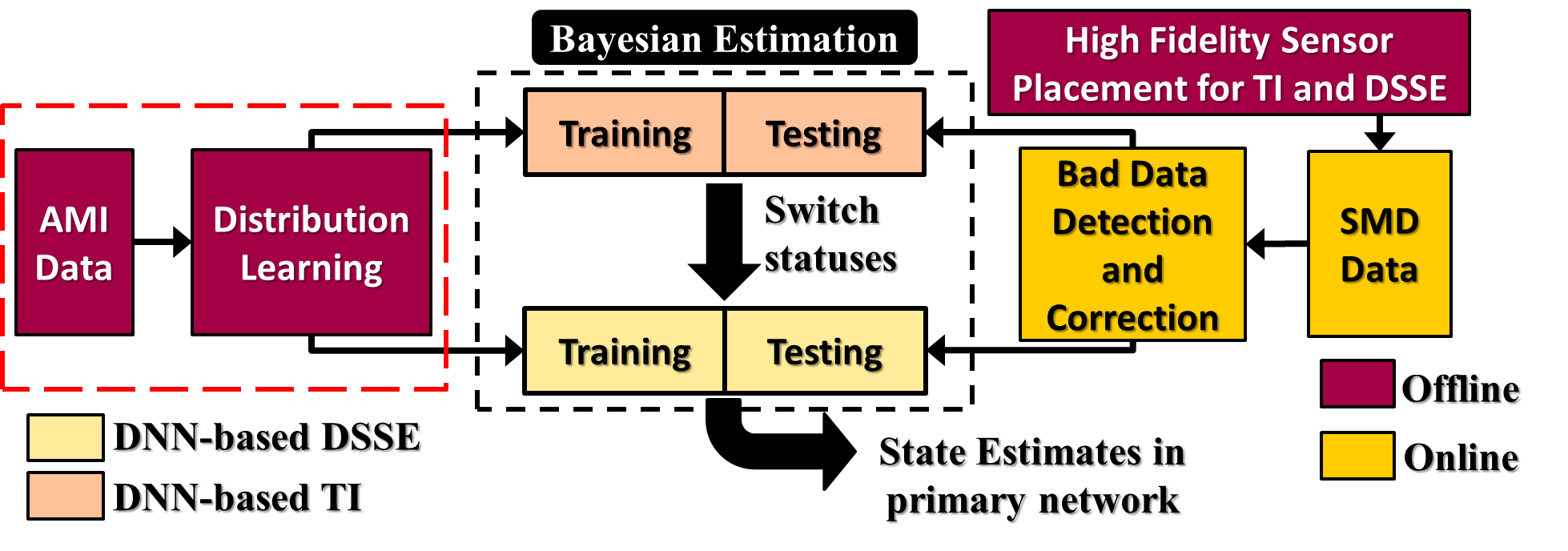}
        \vspace{-2em}
    \caption{Proposed framework overview}
    \label{fig:framework}
    \vspace{-0.5em}
\end{figure}

The overall framework is illustrated in Fig.~\ref{fig:framework}. In the offline phase, historical smart meter data from advanced metering infrastructure (AMI) is used to learn distributions that characterize the aggregated loads at the primary network-level.
In parallel, a correlation-driven SMD placement algorithm identifies the 
required sensor locations for joint TI and DSSE. In the online phase, real-time PMU/$\mu$PMU measurements are fed into the DNN-based TI model, which continuously monitors switch statuses and detects topology changes. Simultaneously, the DNN-based DSSE models produce three-phase voltage phasor estimates for all the nodes. If the DNN-TI detects a topology change, then TL is used to update the DNN-DSSE to adapt it to the new network configuration.

\section{Temporally-Aware Correlation-Driven SMD Placement}
\label{sec:placement}

Time-synchronized
joint TI and DSSE in an incompletely observed distribution network requires SMD locations that simultaneously capture discriminative switching information and are spatially representative of the full network voltage profile. These objectives place competing demands on sensor placement considering TI favors locations sensitive to current redistribution under switching events, while DSSE favors locations that maximize spatial voltage coverage. A placement optimized for either objective alone will underperform on the other task. 
Furthermore, 
distribution networks with significant DER penetration operate under substantially different loading and generation conditions across different hours of the day.
A sensor placement strategy derived from a single hour (as done in \cite{azimian2022})
may fail to provide adequate coverage during other hours.
The proposed algorithm addresses these concerns
by integrating temporal correlation analysis with
spatial correlation-driven SMD placement to satisfy TI accuracy and DSSE performance thresholds simultaneously for all primary circuit voltages across multiple hours.

\subsection{Temporal Correlation Analysis}
\label{subsec:temporal}

\begin{figure}[ht]
    \centering
    \vspace{-0.5em}
    \includegraphics[width=0.38\textwidth]{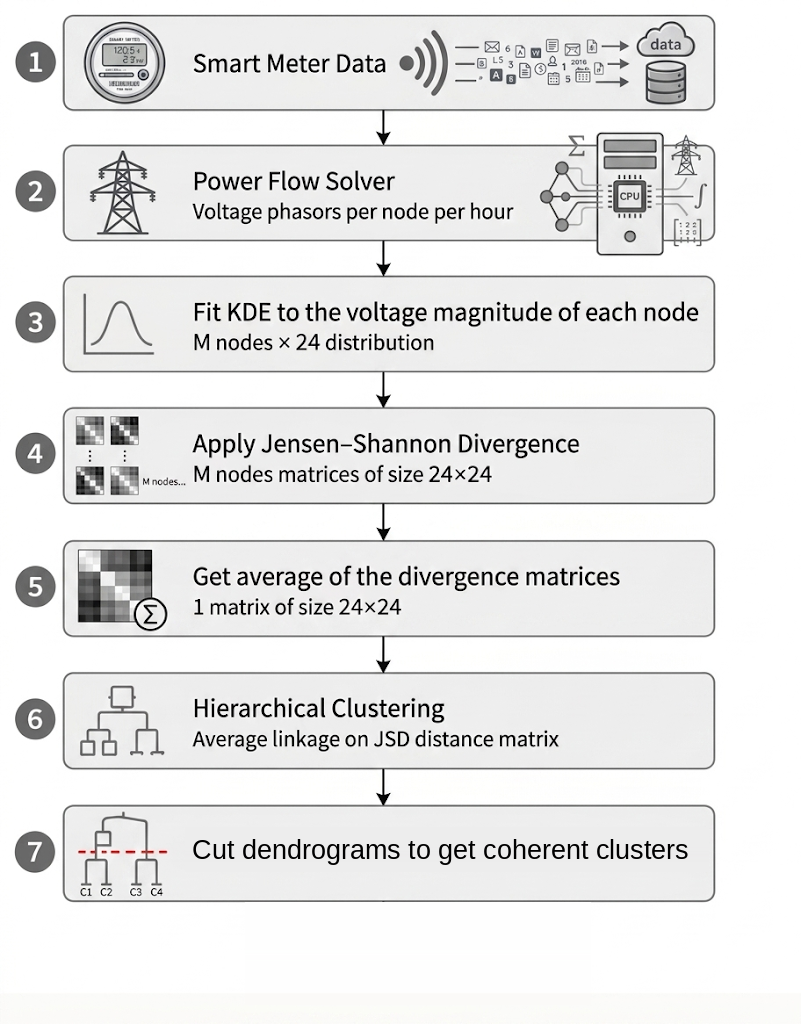}
    \vspace{-3.5em}
    \caption{Temporal correlation analysis pipeline}
    \label{fig:temporal}
    \vspace{-0.5em}
\end{figure}

The goal of this stage is to identify a 
set of $S$ representative time slots whose collective voltage statistics span the full diversity of diurnal
operating conditions, ensuring that the subsequent sensor placement is valid for
all hours of the day.
The procedure is illustrated in
Fig.~\ref{fig:temporal}, and elaborated below.

Historical smart meter data provides hourly net active and reactive power injections at each primary load node. 
Using these injections, large numbers of power flows are
solved
to produce time-series voltage phasor profiles across all network nodes (denoted by $M$). A kernel density estimator (KDE) is then fitted to the voltage magnitude at each node for each hour $h\in\{0,\ldots,23\}$, capturing the non-Gaussian, multimodal character of residential load distributions without imposing a parametric
form \cite{parzen1962}. Next, the Jensen-Shannon divergence (JSD) is computed between the voltage magnitude distributions of every pair of hours $(h_i, h_j)$ at each node $k$, yielding a $24 \times 24$ pairwise dissimilarity matrix $\mathbf{D}^k$. 
Finally, the $\mathbf{D}^k$
are averaged across all the nodes to obtain a 
system-wide dissimilarity matrix, $\bar{\mathbf{D}}$:
\begin{equation}
    \bar{\mathbf{D}} = \frac{1}{M}\sum_{k=1}^{M}\mathbf{D}^k
    \label{eq:jsd_avg}
\end{equation}

Next, hierarchical clustering with average linkage is 
applied to $\bar{\mathbf{D}}$.
The resulting dendrogram is cut where the inter-cluster distance increases most sharply
The threshold of 0.5 is selected at the largest dendrogram merge gap, yielding $S=2$; a lower threshold of 0.4 would produce a third cluster of only 2 hours, which is insufficient to capture the diversity of that regime.

\subsection{
SMD Placement for TI and DSSE}
\label{subsec:placement_alg}

The second stage performs SMD placement using the $S$ representative time slots as one of its inputs.
The other inputs are the
sensor budget, a TI accuracy threshold $\text{TI}_{\text{acc}}$, a DSSE spatial correlation threshold $\text{DSSE}_{\text{corr}}$, and the number of nodes $M$, 
while the output is the 
required SMD location set
$\mathcal{L}$. The complete procedure is summarized
in Algorithm~\ref{alg:placement}.

\subsubsection{Part A: Placement for TI}
Part~A identifies the minimum SMD set whose current phasor measurements enable the DNN-TI model to achieve accuracy $\geq \text{TI}_{\text{acc}}$ across all $S$ time slots. This is formulated as a \textit{sequential forward selection} (SFS) problem, where locations are added greedily to maximize classification accuracy. Current phasors are used as the discriminative feature because topology changes alter the network admittance matrix and
redistribute power flows, producing dominant changes in branch current magnitudes and angles at locations electrically close to the switching event.

\subsubsection{Part B: Placement for DSSE}
Part B augments $\mathcal{L}$ with additional SMDs to ensure that every unmonitored node has a Spearman correlation coefficient (SCC) of at least $\text{DSSE}_{\text{corr}}$ with 
at least one node already in $\mathcal{L}$.
The SCC, computed as shown in \eqref{eq:scc}, is evaluated per feeder/voltage-regulated zone:
\begin{equation}
    \text{SCC}_{kl}^{j} = \rho_s\!\left(V_{j}^{k},V_{j}^{l}\right),
    \; \forall\, k,l \in \{1,\ldots,M\},\; j \in \{\text{mag, ang}\}
    \label{eq:scc}
\end{equation}

This iterative coupling ensures that both objectives are jointly satisfied across all $S$ slots once the TI and DSSE thresholds are met.

\begin{algorithm}[t]
\caption{SMD Placement for DNN-TI and -DSSE}
\label{alg:placement}
\small
\textbf{Inputs:} Budget, $\text{TI}_{\text{acc}}$, $\text{DSSE}_{\text{corr}}$, $M$, $S$\\
\textbf{Output:} Required
SMD locations $\mathcal{L}$\\
\BlankLine
Identify $S$ representative slots via JSD-based hierarchical clustering using the temporal correlation analysis pipeline\\
\BlankLine
\textbf{A. SMD Placement for TI:}\\
A.i \quad Slot $= 1$; \textbf{while} Slot $\leq S$\\
A.ii \quad $N_{\text{feature}} = 1$\\
A.iii \quad If no switches exist, go to (B)\\
A.iv \quad Apply SFS with $N_{\text{feature}}$ features; update $\mathcal{L}$\\
A.v \quad If cost $\geq$ Budget \textit{or} $\text{TI}_{\text{acc}}$ met, go to (B); else $N_{\text{feature}} = N_{\text{feature}} + 1$, go to (A.iv)\\
\BlankLine
\textbf{B. SMD Placement for DSSE:}\\
B.i \quad Compute SCC as per \eqref{eq:scc} for all nodes, phases, and phasor components\\
B.ii \quad If SCC $\geq \text{DSSE}_{\text{corr}}$ $\forall\, k,l,j$, go to (B.vi)\\
B.iii \quad Apply hierarchical clustering to SCC matrix\\
B.iv \quad Form sub-cluster of nodes below $\text{DSSE}_{\text{corr}}$\\
B.v \quad Add highest avg intra-sub-cluster SCC node to $\mathcal{L}$; recompute; if met go to (B.vi), else go to (B.iii)\\
B.vi \quad $\text{Slot} = \text{Slot} + 1$\\
B.vii \quad If $\text{TI}_{\text{acc}}$ satisfied for new slot, go to (B), else go to (A.iv)\\
\end{algorithm}

\section{Dual-Model Deep Inference for Topology and Voltage Phasor Estimation}
\label{sec:dnn}

\begin{figure}[t]
    \centering
    \includegraphics[width=\linewidth]{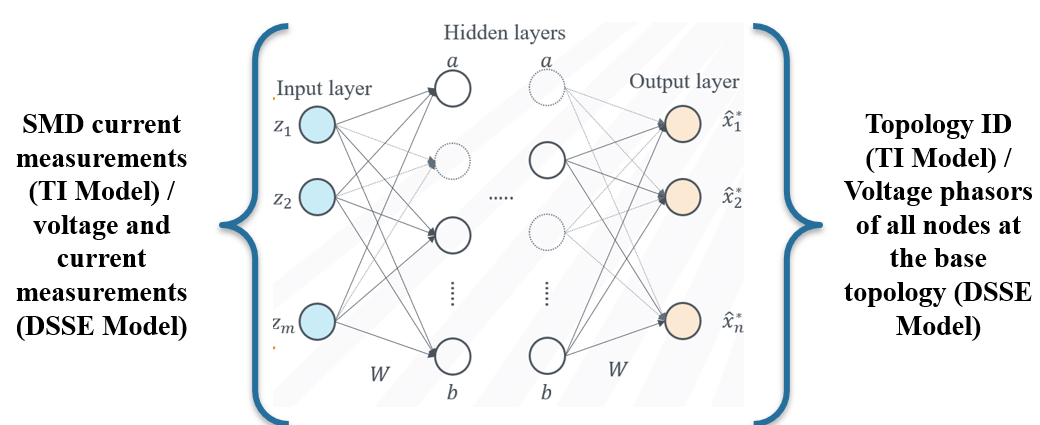}
    \vspace{-1.75em}
    \caption{General DNN architecture used for both TI and DSSE models}
    \label{fig:dnn_architecture}
    \vspace{-1.25em}
\end{figure}

In the proposed framework, TI and DSSE are handled by separate models rather than a single multi-task network. TI requires classification over a discrete topology space while DSSE requires regression over a continuous high-dimensional voltage phasor space. Jointly optimizing both under a shared objective constrains
the expressivity available to each task; separate models allow each network to fully specialize its learned representations. The general DNN architecture shared across all models is illustrated in Fig.~\ref{fig:dnn_architecture}, where $\mathbf{z} = [z_1, \ldots, z_m]^\top$ denotes the SMD measurement input vector and $\hat{\mathbf{x}}^* = [\hat{x}_1^*, \ldots, \hat{x}_n^*]^\top$
denotes the output, representing either topology class probabilities (for the TI model) or estimated voltage phasors (for the DSSE model). 

\subsection{DNN-Based Topology Identification (TI)}
\label{subsec:ti}

The TI model takes as input the three-phase current phasor measurements (magnitude and angle) from all SMD locations in $\mathcal{L}$, and
performs multi-class classification over all feasible network topologies, where each topology corresponds to a unique combination of switch statuses and is assigned a unique integer label. This formulation naturally
handles multiple simultaneous switching operations and scales to large numbers 
of switches \cite{azimian2022}. The model is a fully connected feedforward DNN with hidden layers regularized by batch normalization and dropout. Batch normalization stabilizes training by reducing internal covariate shift, while dropout mitigates overfitting by stochastically deactivating neurons
during training. 
Hyperparameters are tuned using the Keras Tuner framework and reported in Section~\ref{sec:results}.

\subsection{Dual DNN-Based State Estimation}
\label{subsec:dsse}

The DSSE models take as input both voltage and current phasor measurements from the SMD locations in $\mathcal{L}$. Voltage measurements provide direct local observability at instrumented nodes and serve as anchor points from which the network-wide voltage profile is inferred, while current measurements carry the power
flow information necessary to reconstruct unmonitored node
voltages. \textit{Two} 
regression models are trained independently in parallel, one for three-phase voltage magnitudes and the other for three-phase voltage angles, across all $M$ 
nodes. This dual-model design is motivated by the observation that voltage magnitudes and angles exhibit fundamentally different spatial correlation structures and dynamic ranges across the network. A single unified model must learn both mappings simultaneously under a shared loss, constraining the internal representations available to each task. Separate models allow full specialization and, as demonstrated in Section~\ref{sec:results}, consistently produce lower estimation error than the unified approach, particularly when evaluated across multiple operating time slots where the statistical characteristics of magnitude and angle
distributions diverge.

Each model is a fully connected feedforward DNN with hidden layers regularized by batch normalization and dropout. The number of hidden layers, neurons per layer, and dropout rate are determined independently for each model and each representative time slot using the Keras Tuner framework.
The specific architectures 
for each slot are reported in Section~\ref{sec:results}.


\section{Topology-Triggered Model Adaptation Through Fine-Tuning (FT)}
\label{sec:tl}
The dual DNN-DSSE models developed in Section~\ref{sec:dnn} are trained on data generated from the
base topology. When the network undergoes a switching operation, the marginal distribution of SMD measurements shifts because loads are now served through different paths and branch currents are
redistributed accordingly. A model trained on the preceding topology will receive test inputs from a shifted distribution, degrading estimation accuracy in proportion to the structural difference between topologies. Full retraining from scratch for every new topology is theoretically sound but practically
infeasible, requiring large amounts of new training data and significant computational time.

TL
addresses this issue by reusing knowledge encoded in a previously trained model to accelerate learning in a new but related domain \cite{pan2010}. In the context of DNN-based DSSE, the source domain corresponds to the topology under which the model was most recently trained, and the target domain corresponds to the new topology after a switching event.
The feature space does not change across domains since the same SMD measurements are used regardless of topology; however, the marginal distribution of those measurements does change, making adaptation necessary. Among available TL strategies, parameter-based transfer via FT
is adopted. FT
initializes the target-domain model with the weights of
the most recently trained model and continues training on a limited amount of target-domain data, rather than from random initialization. 
The rationale for using FT is that distribution networks sharing the same physical infrastructure across topologies exhibit strongly overlapping voltage and current statistics.
Hence, lower DNN layers 
learn
general representations of the network's electrical behavior, such as feeder impedance profiles and load correlation structures, that remain approximately valid after switching. Only the higher layers encoding topology-specific mappings require 
updating, making fine-tuning from existing weights more data-efficient than retraining from scratch. 

The proposed TL framework adopts a strictly sequential adaptation strategy: whenever a topology change is detected by the DNN-TI model, the active DSSE model is fine-tuned directly from its current weights.
This imposes minimal memory requirements while remaining 
adaptive to the evolving network configuration. Formally, let $\boldsymbol{\theta}^{(k)}$ denote the DSSE model weights after the $k$-th topology transition, trained on topology $\mathcal{T}_k$. When the network transitions to $\mathcal{T}_{k+1}$, a small dataset $\mathcal{D}_{k+1} = \{(\mathbf{z}_i, \mathbf{x}_i)\}_{i=1}^{N_t}$ is used for the new topology, where $N_t \ll N_s$ (the original training set size), with $N_t$ chosen as the minimum number of scenarios at which post-adaptation DSSE accuracy stabilizes.
The adapted model is obtained by minimizing the empirical mean-squared error (MSE) loss initialized from the current weights:
\begin{equation}
    \boldsymbol{\theta}^{(k+1)} = \arg\min_{\boldsymbol{\theta}}
    \frac{1}{N_t} \sum_{i=1}^{N_t}
    \left\| \mathbf{x}_i - K\!\left(\mathbf{z}_i;\, \boldsymbol{\theta}
    \right) \right\|^2, \quad
    \boldsymbol{\theta}_0 = \boldsymbol{\theta}^{(k)}
    \label{eq:finetune}
\end{equation}
where $K(\cdot\,;\boldsymbol{\theta})$ is the DNN mapping. The updated model $\boldsymbol{\theta}^{(k+1)}$ replaces the active model and is used for all subsequent inference until the next topology change is detected.

The complete sequential adaptation procedure is summarized in
Algorithm~\ref{alg:tl}. During real-time operation, the DNN-TI model continuously monitors incoming SMD measurements and classifies the current topology. Upon detecting a change, FT is triggered using $N_t$ pre-generated and stored scenarios for the detected topology, and the DSSE models resume 
inference
post-update.
The results obtained using the proposed topology-triggered model adaptation are provided in Section~\ref{sec:results} for four consecutive topology transitions.

\begin{algorithm}[ht]
\caption{Topology-Triggered Sequential FT}
\label{alg:tl}
\small
\textbf{Inputs:} SMD measurements $\mathbf{z}$, current model
$\boldsymbol{\theta}^{(k)}$, DNN-TI model, current topology $\mathcal{T}_k$\\
\textbf{Output:} State estimate $\hat{\mathbf{x}}$; updated model
$\boldsymbol{\theta}^{(k+1)}$ if topology change occurs\\
\BlankLine
1. Feed $\mathbf{z}$ into DNN-TI; obtain predicted topology $\hat{\mathcal{T}}$\\
2. If $\hat{\mathcal{T}} = \mathcal{T}_k$: compute
$\hat{\mathbf{x}} = K(\mathbf{z};\,\boldsymbol{\theta}^{(k)})$; return
$\hat{\mathbf{x}}$\\
3. If $\hat{\mathcal{T}} \neq \mathcal{T}_k$: set
$\mathcal{T}_{k+1} = \hat{\mathcal{T}}$\\
4. Generate $N_t$ scenarios for $\mathcal{T}_{k+1}$; minimize
\eqref{eq:finetune} from $\boldsymbol{\theta}^{(k)}$; obtain
$\boldsymbol{\theta}^{(k+1)}$\\
5. Set $\boldsymbol{\theta}^{(k)} \leftarrow \boldsymbol{\theta}^{(k+1)}$,
$\mathcal{T}_k \leftarrow \mathcal{T}_{k+1}$\\
6. Compute $\hat{\mathbf{x}} = K(\mathbf{z};\,\boldsymbol{\theta}^{(k)})$;
return $\hat{\mathbf{x}}$\\
\end{algorithm}

\section{Simulation Results}
\label{sec:results}

\subsection{Test System and Data Generation}

Simulations are conducted on a 
240-node US Midwest primary distribution test system that is described in \cite{bu2019}. The system 
has three feeders, and 9 sectionalizing switches yielding 84 feasible 
topologies. 
Of the 240 nodes of this system, 193 nodes 
have loads.
One year of hourly smart meter data 
is publicly available for these loads.
To generate training datasets of sufficient sizes for DNN training, a KDE is fitted
independently to the active and reactive power injection at each load node per hour obtained from the smart meter data.
Monte Carlo sampling is then done to draw new load scenarios from the learned distributions. 
Power flow is solved in OpenDSS \cite{opendss} for each sampled scenario to obtain the corresponding voltage and current phasors. 
For the base-topology DNN-DSSE, 1,500 scenarios per hour are generated (36,000 total), and for DNN-TI, 100 scenarios per hour per topology are used. For FT,
150 scenarios per hour (3,600 total) are generated for each target topology. Three noise models are applied to SMD measurements to evaluate robustness, namely,
Gaussian $\mathcal{N}(0,\sigma^2)$, Gaussian mixture model (GMM) as a
two-component mixture $\sum_{c=1}^{2}\pi_c\,\mathcal{N}(\mu_c,\sigma_c^2)$, and Laplacian $\mathrm{Lap}(0,b)$. The last two represent non-Gaussian characteristics consistent with field SMD measurements \cite{azimian2020}.
All simulations are conducted on a server equipped with dual Intel Xeon Platinum 8368 processors (76 cores, 152 logical processors at 2.40\,GHz) and 384\,GB of DDR4-3200 RAM.

\subsection{Proposed DNN-Based TI and DSSE}

\begin{figure}[t]
    \centering
    \includegraphics[width=0.92\linewidth]{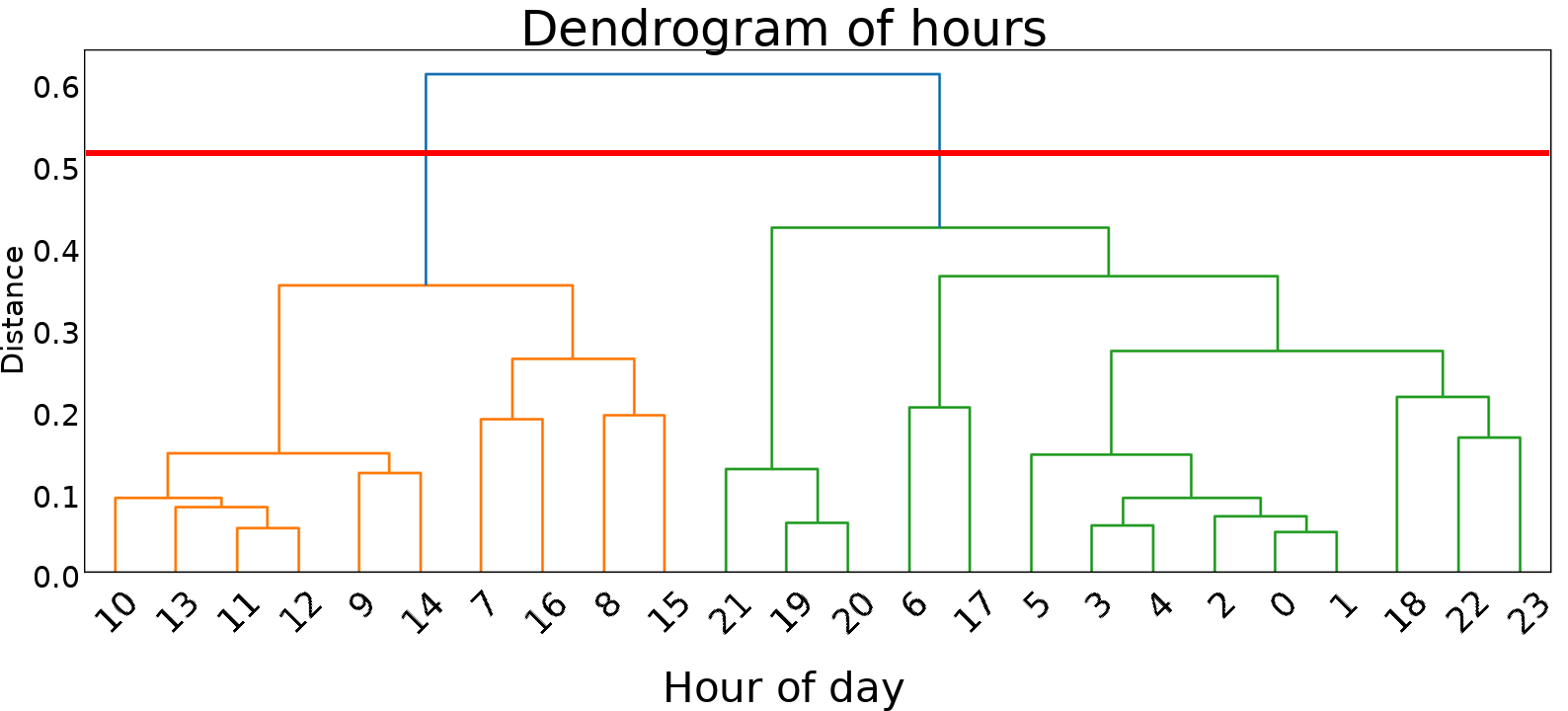}
    \vspace{-0.1in}
    \caption{Temporal correlation results for JSD-based hierarchical clustering}
    \vspace{-0.2in}
    \label{fig:dendrogram}
\end{figure}

The JSD-based hierarchical clustering partitions the 24 hours of the day into $S=2$ representative time slots at a distance threshold of 0.5, as shown in the dendrogram of Fig.~\ref{fig:dendrogram}. Slot~1 captures daytime and evening hours (7--22) with higher voltage variability, 
while Slot~2 corresponds to overnight hours (22--7) with lower voltage
variability. Applying Algorithm~\ref{alg:placement} with $\text{TI}_{\text{acc}}=99\%$ and $\text{DSSE}_{\text{corr}}=0.98$,
yields 7 SMD locations after Slot~1 and one additional location
after incorporating Slot~2, resulting in a total of 8 locations for placing SMDs.

\begin{figure*}[t]
    \centering
    \includegraphics[width=0.85\textwidth,height=2.4in]{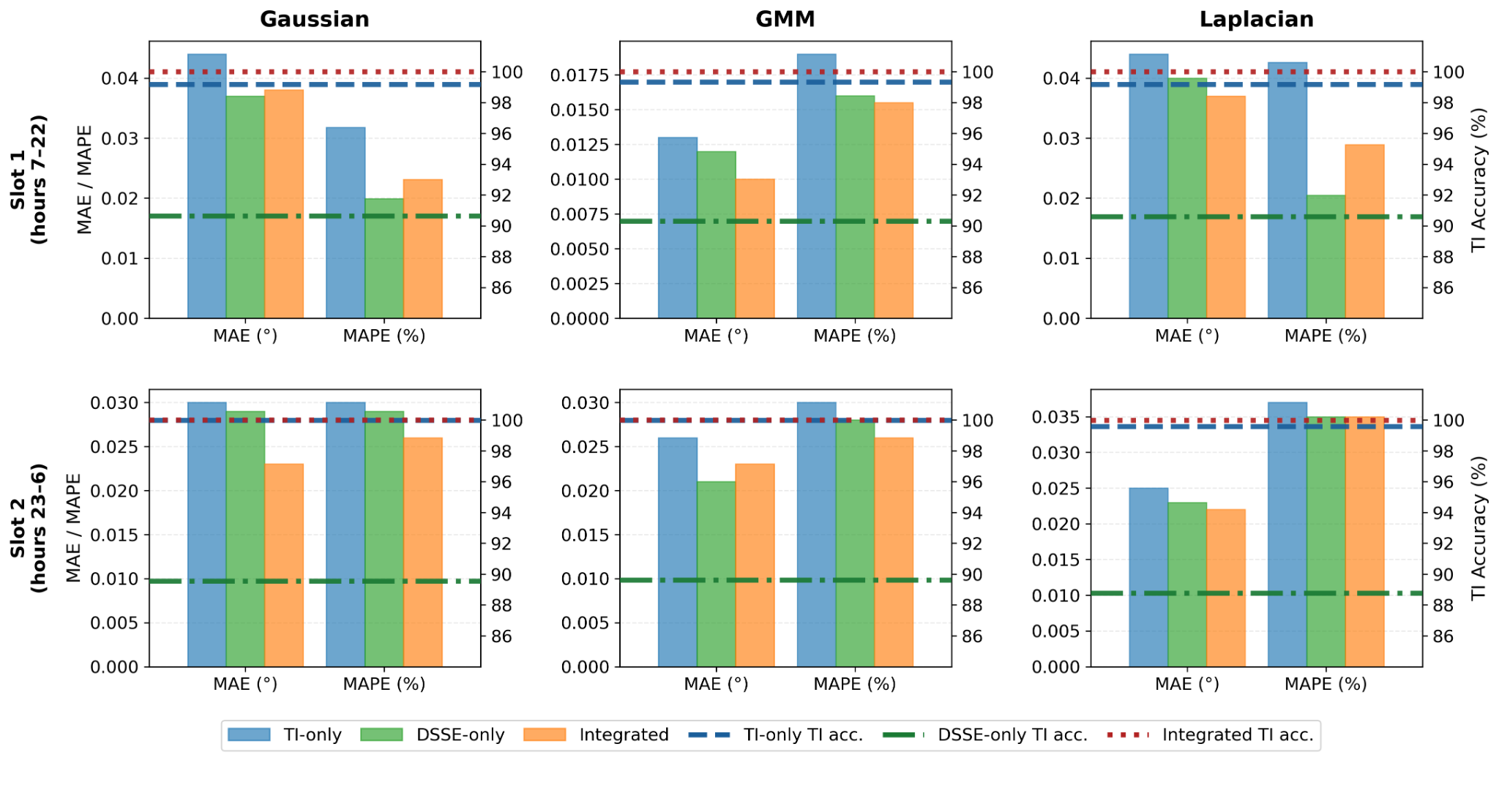}
        \vspace{-0.2in}
    \caption{Robustness of proposed placement approach under different noises}
    \label{fig:placement_bar}
    \vspace{-0.2in}
\end{figure*}

Fig.~\ref{fig:placement_bar} compares the three SMD placement strategies, namely, TI-only, DSSE-only, and the proposed integrated placement, under all three noise conditions and for both time slots. The DSSE-only placement consistently degrades TI accuracy to the 88--91\% range, falling below the 99\% operational threshold regardless of noise type or time slot. The TI-only placement maintains high TI accuracy but fails to meet DSSE spatial
coverage requirements. The integrated placement satisfies both objectives simultaneously across all noise types and 
time slots, achieving TI accuracy above 99.9\%, magnitude MAPE within 0.04\%, and angle MAE within 0.03$^\circ$. The SMD locations for the integrated placement and corresponding performance metrics are reported in the last row
of Table~\ref{tab:robustness}.

\begin{table}[ht]
\vspace{-0.12in}
\centering
\caption{Comparing One Unified Model to Dual Model DNN-DSSE Under GMM Noise}
\vspace{-0.5em}
\label{tab:robustness}
\footnotesize
\setlength{\tabcolsep}{2pt}
\begin{tabular}{l|c|c|c}
\hline
\textbf{Model} & \textbf{Slot} & \textbf{Phase MAE ($^\circ$)} &
\textbf{Mag. MAPE (\%)} \\
\hline
\multirow{2}{*}{Single
Unified
DNN}       & Slot 1 & 0.019 & 0.031 \\
\cline{2-4}
                                    & Slot 2 & 0.044 & 0.062 \\
\hline
\multirow{2}{*}{Dual DNN (proposed)} & Slot 1 & 0.010 & 0.015 \\
\cline{2-4}
                                      & Slot 2 & 0.023 & 0.026 \\
\hline
\multicolumn{4}{p{0.88\columnwidth}}{\footnotesize \textbf{SMD locations:}
1--1001, 1008--1009, 2011--2012, 2019--2021, 3003--3005, 3068--3075,
3081--2016 (inter-feeder B--C link), 3137--3138} \\
\hline
\end{tabular}
\end{table}

Table~\ref{tab:robustness} benchmarks the proposed dual-model design against a single 
unified 
DNN estimating both magnitude and angle simultaneously under GMM noise for both time slots. 
The proposed dual model reduces Mag.~MAPE by 52\% and Ang.~MAE by 47\% in Slot~1, and by 58\% and 48\%, respectively, in Slot~2, confirming that
specialized models consistently outperform a single
unified 
estimator across both operating regimes. The larger degradation of the single
unified 
model in Slot~2 further confirms that
when trained for
one representative hour, the single model 
struggles to generalize across other
operating envelopes.

The DNN architectures for the base topology under GMM noise are summarized in Table~\ref{tab:hyper}. The TI model uses a deeper 6-layer network with 800 neurons per layer to resolve the 84-class topology space, while the DSSE magnitude and angle models use shallower architectures independently tuned by Keras Tuner. The fine-tuned model inherits the base 
architecture with a reduced learning rate of $10^{-4}$ and only 50 epochs, enabling rapid adaptation upon topology change.

\begin{table}[t]
\centering
\caption{Hyperparameters for DNN-Based TI and DSSE 
}
\vspace{-0.5em}
\label{tab:hyper}
\small
\setlength{\tabcolsep}{3pt}
\begin{tabular}{l|c|c|c|c}
\hline
\multirow{2}{*}{\textbf{Parameter}} & \multirow{2}{*}{\textbf{DNN-TI}}
& \multicolumn{2}{c|}{\textbf{DNN-DSSE}}
& \multirow{2}{*}{\textbf{FT}} \\
\cline{3-4}
& & \textbf{Mag} & \textbf{Ang} & \\
\hline
Inputs         & 48 (I)    & \multicolumn{3}{c}{92 (V\&I of SMDs)} \\
\hline
Outputs        & 84 (topo) & \multicolumn{3}{c}{462 (states)} \\
\hline
Hidden layers  & 6         & \multicolumn{2}{c|}{2}                 & \\
\hline
Neurons/layer  & 800       & 640        & 256                       & \\
\hline
Dropout        & 0.3       & \multicolumn{2}{c|}{0.5}               & \\
\hline
Hidden activ.  & \multicolumn{4}{c}{ReLU} \\
\hline
Output activ.  & Softmax   & \multicolumn{3}{c}{Linear} \\
\hline
Optimizer      & \multicolumn{4}{c}{Adam} \\
\hline
Loss           & \makecell{Categorical \\ cross-entropy} & \multicolumn{3}{c}{MSE} \\
\hline
Epochs         & 100       & \multicolumn{2}{c|}{200}               & 50 \\
\hline
Batch size     & 105       & \multicolumn{2}{c|}{32}                & 32 \\
\hline
Learning rate  & $10^{-3}$ & \multicolumn{2}{c|}{$10^{-3}$}        & $10^{-4}$ \\
\hline
Samples/hour   & 8400      & \multicolumn{2}{c|}{1500}              & 150 \\
\hline
\end{tabular}
\end{table}

Table~\ref{tab:lse} compares the proposed DNN-based approach to
linear state estimation (LSE) under Gaussian and non-Gaussian (GMM) noises for both time slots. LSE requires full observability and is implemented
with 113 SMD locations (determined using the approach developed in \cite{biswas2020}), while the proposed DNN operates with only 8 SMDs, about a 93\% reduction in instrumentation needs. 
The DNN achieves slightly worse results than
LSE (particularly in the magnitudes), while it
has similar results under Gaussian and GMM noises, demonstrating its 
robustness 
to different noise models.

\begin{table}[ht]
\vspace{-0.1in}
\centering
\caption{Proposed DNN vs.\ Conventional LSE}
\vspace{-0.5em}
\label{tab:lse}
\small
\setlength{\tabcolsep}{3pt}
\begin{tabular}{l|c|c|c|c|c}
\hline
\multirow{2}{*}{\textbf{Model}} & \multicolumn{2}{c|}{\textbf{Slot 1}} &
\multicolumn{2}{c|}{\textbf{Slot 2}} & \multirow{2}{*}{\textbf{\#SMDs}} \\
\cline{2-5}
& \textbf{MAE} & \textbf{MAPE} & \textbf{MAE} &
\textbf{MAPE} & \\
\hline
LSE (Gaussian)     & 0.001$^\circ$ & 0.015\% & 0.001$^\circ$ & 0.014\% & 113 \\
\hline
DNN (Gaussian)     & 0.017$^\circ$ & 0.018\% & 0.024$^\circ$ & 0.026\% & 8   \\
\hline
DNN (GMM) & 0.010$^\circ$ & 0.015\% & 0.028$^\circ$ & 0.064\% & 8   \\
\hline
\end{tabular}
\end{table}

\subsection{TL for Topology Adaptation}

\begin{figure}[ht]
    \vspace{-0.1in}
    \centering
    \includegraphics[width=1\linewidth,trim=0.15in 0.15in 0.15in 0.10in,clip]{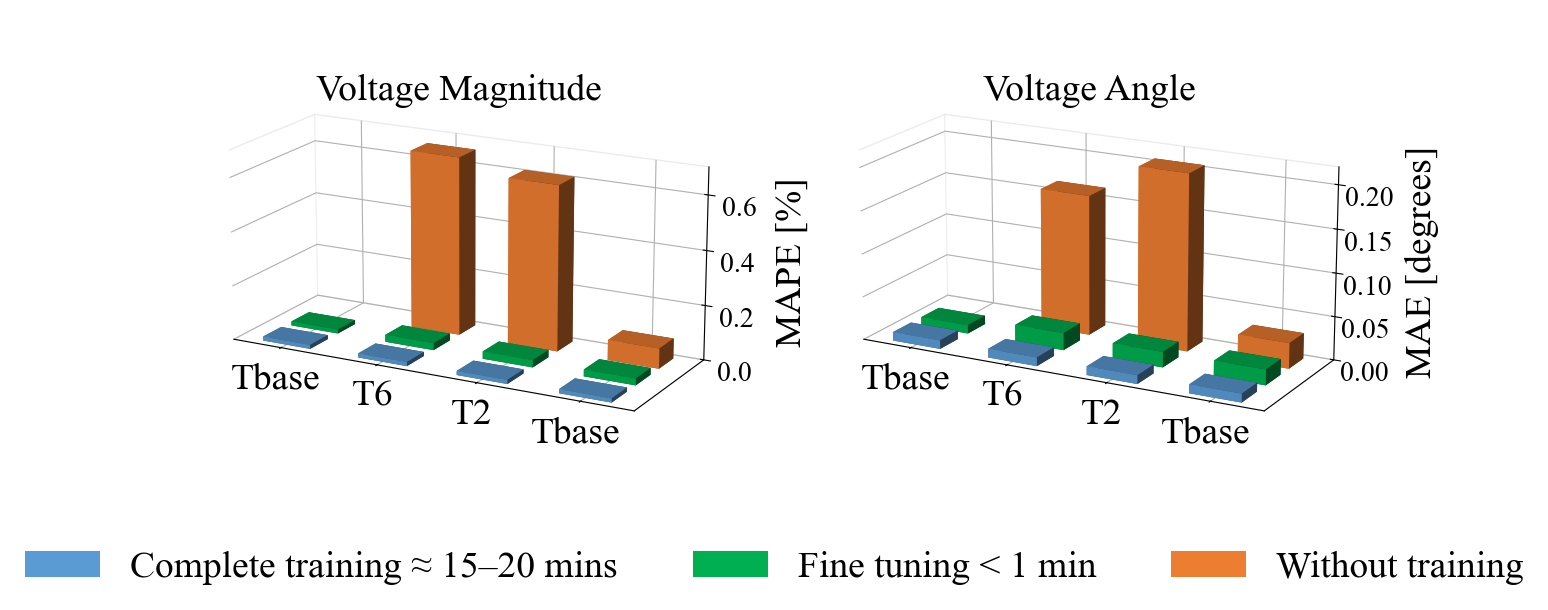}
    \vspace{-0.2in}
    \caption{Topology adaptation: Voltage magnitude MAPE \& Voltage angle MAE}
    \label{fig:Topology Adaptation}
    \vspace{-0.1in}
\end{figure}

Fig.~\ref{fig:Topology Adaptation} shows MAPE and MAE
across the topology transition sequence under GMM noise for Slot~1.
Table~\ref{tab:tl} reports the exact metrics and training times for the sequence:
$\mathrm{Tbase}\rightarrow \mathrm{T36} \rightarrow \mathrm{T2}\rightarrow \mathrm{Tbase}$, 
for three conditions, namely,
no adaptation, FT,
and complete retraining. $\mathrm{T36}$ differs from $\mathrm{Tbase}$ in five switch positions and $\mathrm{T2}$ in two switch positions, exercising both large and moderate topological shifts. Without adaptation, MAPE degrades to 0.072--0.090\% and angle MAE reaches up to 0.098$^\circ$. 
FT helps maintain accuracies of 0.025--0.027\% MAPE and
0.018--0.020$^\circ$ MAE, which approaches 
complete-retraining accuracy at a fraction of the computational cost/time. The return to $\mathrm{Tbase}$ demonstrates that sequential FT from an intermediate model state is sufficient for accurate recovery even when
revisiting a previously encountered topology, without storing any model checkpoints. Exact metrics for all transitions are reported in Table~\ref{tab:tl}.

\begin{table}[ht]
\centering
\caption{FT Results Across Topology Transitions (GMM, Slot~1)}
\vspace{-0.5em}
\label{tab:tl}
\small
\setlength{\tabcolsep}{2.5pt}
\begin{tabular}{l|c|c|c|c}
\hline
\textbf{Metric} & \textbf{Tbase} & \textbf{Tbase$\rightarrow$T36} &
\textbf{T36$\rightarrow$T2} & \textbf{T2$\rightarrow$Tbase} \\
\hline
MAPE (\%)              & 0.015 & 0.026 & 0.027 & 0.025 \\
\hline
MAE ($^\circ$)         & 0.010 & 0.020 & 0.018 & 0.018 \\
\hline
Mag.\ train time (s)   & 1480  & 31    & 28    & 27    \\
\hline
Ang.\ train time (s)   & 947   & 44    & 43    & 43    \\
\hline
\end{tabular}
\end{table}

\section{Conclusion}
\label{sec:conclusion}

This paper proposed an integrated deep learning framework for simultaneous TI and three-phase unbalanced DSSE in real-time unobservable primary networks instrumented by a minimal set of SMDs. 
The framework is validated using a 240-node real-world  
test system under full 
operational variability.
A correlation-driven placement algorithm was introduced first that jointly
optimized sensor locations for both TI and DSSE by exploiting temporal variability across different hours
via JSD-based hierarchical clustering and spatial correlation of nodal voltage phasors via 
SCC-based rank
analysis. 
For the identified test system,
the algorithm identified two representative time
slots and converged to 8 SMD locations satisfying both a $99\%$ TI accuracy threshold and a $0.98$ SCC threshold simultaneously across all noise conditions, compared to 113 SMD locations required by conventional LSE.

Dedicated dual DNN-DSSE models for voltage magnitude and angle were developed next and it was shown that they outperformed
a single 
unified 
estimator by up to $52\%$ in magnitude MAPE and $48\%$ in angle MAE under GMM noise, with consistent gains across both time slots. The framework demonstrated robustness under Gaussian, GMM, and Laplacian noises, maintaining TI accuracy above $99.9\%$ and DSSE errors within operational bounds in all cases. Finally, FT-based TL enabled rapid topology adaptation upon switching events, resulting in
an improvement of $67$-$71\%$ in MAPE and $37$-$80\%$ in MAE over the no adaptation case
across three consecutive topology transitions. The adaptation took place in under one minute without storing any model checkpoints,
compared to $15$-$25$ minutes for complete retraining.

\end{document}